\documentclass[aps,prd,preprint,showpacs,showkeys,superscriptaddress,nofootinbib]{revtex4-1}

\usepackage{booktabs}
\usepackage{graphicx}
\usepackage{xcolor}
\usepackage{amsmath}
\usepackage{hyperref}
\usepackage{bm}
\usepackage{mathtools}
\usepackage{amssymb}
\usepackage{lipsum}
\usepackage{epstopdf}
\usepackage{slashed}
\usepackage{multirow}
\usepackage{appendix}
\usepackage{cleveref}

\AtBeginDocument{
\heavyrulewidth=.08em
\lightrulewidth=.05em
\cmidrulewidth=.03em
\belowrulesep=.65ex
\belowbottomsep=0pt
\aboverulesep=.4ex
\abovetopsep=0pt
\cmidrulesep=\doublerulesep
\cmidrulekern=.5em
\defaultaddspace=.5em
}

\allowdisplaybreaks

\begin{document}
\def\qq{\langle \bar q q \rangle}
\def\uu{\langle \bar u u \rangle}
\def\dd{\langle \bar d d \rangle}
\def\sp{\langle \bar s s \rangle}
\def\GG{\langle g_s^2 GG \rangle}
\def\Tr{\mbox{Tr}}

\def\lrar{\leftrightarrow}  

\def\ds{\displaystyle}
\def\beq{\begin{equation}}
\def\eeq{\end{equation}}
\def\bea{\begin{eqnarray}}
\def\eea{\end{eqnarray}}
\def\beeq{\begin{eqnarray}}
\def\eeeq{\end{eqnarray}}
\def\ve{\vert}
\def\vel{\left|}
\def\ver{\right|}
\def\nnb{\nonumber}
\def\ga{\left(}
\def\dr{\right)}
\def\aga{\left\{}
\def\adr{\right\}}
\def\lla{\left<}
\def\rra{\right>}
\def\rar{\rightarrow}
\def\nnb{\nonumber}
\def\la{\langle}
\def\ra{\rangle}
\def\ba{\begin{array}}
\def\ea{\end{array}}
\def\tr{\mbox{Tr}}
\def\ssp{{\Sigma^{*+}}}
\def\sso{{\Sigma^{*0}}}
\def\ssm{{\Sigma^{*-}}}
\def\xis0{{\Xi^{*0}}}
\def\xism{{\Xi^{*-}}}

\def\qs{\la \bar s s \ra}
\def\qu{\la \bar u u \ra}
\def\qd{\la \bar d d \ra}

\def\gGgG{\la g^2 G^2 \ra}
\def\q{\gamma_5 \not\!q}
\def\x{\gamma_5 \not\!x}
\def\g5{\gamma_5}
\def\sb{S_Q^{cf}}
\def\sd{S_d^{be}}
\def\su{S_u^{ad}}
\def\sbp{{S}_Q^{'cf}}
\def\sdp{{S}_d^{'be}}
\def\sup{{S}_u^{'ad}}
\def\ssp{{S}_s^{'??}}

\def\sig{\sigma_{\mu \nu} \gamma_5 p^\mu q^\nu}
\def\fo{f_0(\frac{s_0}{M^2})}
\def\ffi{f_1(\frac{s_0}{M^2})}
\def\fii{f_2(\frac{s_0}{M^2})}
\def\O{{\cal O}}
\def\sl{{\Sigma^0 \Lambda}}
\def\es{\!\!\! &~=~& \!\!\!}
\def\ap{\!\!\! &\approx& \!\!\!}
\def\ar{&+& \!\!\!}
\def\ek{&-& \!\!\!}
\def\kek{\!\!\!&-& \!\!\!}
\def\cp{&\times& \!\!\!}
\def\se{\!\!\! &\simeq& \!\!\!}
\def\eqv{&\equiv& \!\!\!}
\def\kpm{&\pm& \!\!\!}
\def\kmp{&\mp& \!\!\!}
\def\mcdot{\!\cdot\!}
\def\erar{&\rightarrow&}


\title{Light-cone sum rules for radial excitation of decuplet to octet baryons electromagnetic transition form factors}

\author{T.~M.~Aliev}
\email{taliev@metu.edu.tr}
\affiliation{Department of Physics, Middle East Technical University, Ankara, 06800, Turkey}


\author{S.~Bilmis}
\email{sbilmis@metu.edu.tr}
\affiliation{Department of Physics, Middle East Technical University, Ankara, 06800, Turkey}
\affiliation{TUBITAK ULAKBIM, Ankara, 06530, Turkey}

\author{M.~Savci}
\email{savci@metu.edu.tr}
\affiliation{Department of Physics, Middle East Technical University, Ankara, 06800, Turkey}

\date{\today}

\begin{abstract}
 Magnetic dipole moment form factor, $G_M^{(2)}(Q^2)$, describing the radial excitation of decuplet baryons to octet baryons electromagnetic transitions as well as the ratios, $R_{SM} = - \frac{1}{4 m_2^2} \sqrt{4 m_2^2 Q^2 + (m_2^2 - Q^2 -m_1^2)^2} \frac{G_C^{(2)}(Q^2)}{G_M^{(2)}(Q^2)}$ and $R_{EM} = -\frac{G_E^{(2)}(Q^2)}{G_M^{(2)}(Q^2)}$ are calculated in the framework of light-cone sum rules. We also estimate the degree of the violation of U-spin symmetry. The obtained results for the multipole form factors can be useful in searching the properties of radially excited baryon states.

\end{abstract}

\maketitle

\section{Introduction}
The considerable progress in studying the spectrum of hadrons and obtaining useful information about their inner structure using experimental data with the electromagnetic interaction has essentially extended our knowledge about the QCD dynamics at low energies. The properties of the excited states essentially depend on the properties of QCD at low energy.

Many baryon states predicted by the quark model have been observed, and their properties (masses, decay widths, quantum numbers) are carefully studied~\cite{PhysRevD.98.030001}. However, part of the baryons  predicted by the quark model have not been discovered in experiments yet. These states are called ``missing-states''~\cite{koniuk1980have}. Searching for these baryons constitutes one of the main motivations for future studies in the physics program of Jefferson Lab, BATE, and MAMI facilities, namely investigation of the spectrum of baryons containing strange quarks. Considering the potential of these facilities, it is expected that these investigations will enrich our knowledge on the application of QCD in this direction, and more information will be acquired about the properties of hadrons around their first excitations via electromagnetic probes.

In literature, there are many studies in this area. For instance, the electroproduction of the first radial excitation of $\Delta$ baryon is investigated in the framework of the light-front model~\cite{Capstick:1994ne}, and the effect of $\Delta(1600)$ is studied in~\cite{Burkert:2019opk}. Moreover, the transition $\gamma^* N \rightarrow \Delta(1232),~ \Delta(1600)$ is analyzed within a diquark-diquark and covariant constituent of the quark model in~\cite{Lu:2019bjs,Ramalho:2010cw}, correspondingly. The transition form factors  $\gamma^* N \rightarrow \Delta(1600)$ within light-cone sum rules is studied in~\cite{Aliev:2020ume}.
In this study, the form factors of the other members of the radial excitations decuplet baryons in reactions $\gamma^* \Sigma \rightarrow \Sigma^{*}$ and $\gamma^* \Xi \rightarrow \Xi^{*}$ are calculated within the light-cone sum rules (LCSR). Here $\gamma^*$ means virtual photon, and $B^{*}$ is the radial excitation of ground-state decuplet baryons. Note that the form factors $\gamma^* N \rightarrow \Delta(1232)$ and $\gamma^* \text{ octet} \rightarrow \text{ decuplet}$ baryon transition within the same approach is studied in~\cite{Braun:2005be} and \cite{Aliev:2013jta}, respectively.

The paper is organized as follows. In section~\ref{sec:2}, we derive the sum rules for the multipole form factors responsible for $\gamma^* B \rightarrow B^{*}$ transition. In section~\ref{sec:3}, the numerical analysis of the derived sum rules for multipole form factors is performed. The summary and conclusion of the study are presented in the last section.
\section{Light-cone sum rules for the $\gamma^* B  \rightarrow B^*$ transition form factors}
\label{sec:2}
In this section, the light-cone sum rules for the $\gamma^* B \rightarrow B^{*} $ transition form factors is obtained. These form factors are defined by the matrix element of the electromagnetic current $j_\mu^{el} = e_{q_1} \bar{q}_1 \gamma_\mu q_1 + e_{q_2} \bar{q}_2 \gamma_\mu q_2  + e_{q_3} \bar{q}_3 \gamma_\mu q_3$ between the initial ground octet and first radial excitations of decuplet baryon states, i.e., $ \langle B^* (p^\prime) | j_\mu^{el} | B(p) \rangle$. In general form, this matrix element can be written as
\begin{equation}
  \label{eq:1}
  \langle  B^*(p^\prime) j_\mu^{el}| B(p) \rangle = \bar{u}_\beta(p^\prime) \Gamma_{\beta \mu}u(p).
\end{equation} 
The main issue is defining constraint-free and gauge-invariant form factors from this matrix element. This problem is solved in~\cite{Jones:1972ky}, and the matrix element in Eq.~\eqref{eq:1} is determined as
\begin{equation}
  \label{eq:2}
  \begin{split}
     \langle B^*(p^\prime) | j_\mu^{el} | B(p) \rangle &= \sum_{i=1}^{2} \bar{u}_\beta(p^\prime) \Bigl\{ G_1^{(i)}(Q^2) (- q_\beta \gamma_\mu + \slashed{q} g_{\beta \mu}) + G_2^{(i)}(Q^2)(-q_\beta \mathcal{P}_\mu + (q \mathcal{P}) g_{\beta \mu}) \\
     &+ G_3^{(i)} (q_\beta q_\mu - q^2 g_{\beta \mu})  \Bigr\}
    \gamma_5 u(p),
  \end{split}
\end{equation}
where $i=1(2)$ corresponds to the ground (first radial excitation) state decuplet baryons, $q= p - p^\prime$, $\mathcal{P}_\alpha = \frac{1}{2}(p + p^\prime)_\alpha$, and $u_{\beta}(p^\prime)$ is the Rarita-Schwinger spinor for spin $3/2$ baryons.

From the experimental point of view, the multipole form factors are more convenient to study. The relations among form factors $G_i(Q^2)$ and multipole magnetic dipole $G_M^{(2)}(Q^2)$, electric quadrupole, $G_E^{(2)}{(Q^2)}$, and Coulomb quadrupole, $G_C^{(2)}(Q^2)$ form factors are obtained in~\cite{Devenish:1975jd,Jones:1972ky}.

\begin{equation}
  \label{eq:3}
  \begin{split}
    G_M^{(i)} (Q^2) &= \frac{m_0}{3(m_0 + m_i)} \big[  (3(m_i + m_0)(m_i + m_0) + Q^2) \frac{G_1^{(i)}(Q^2)}{m_i} \\
    &+ (m_i^2 - m_0^2) G_2^{(i)}(Q^2) - 2Q^2 G_3^{(i)}(Q^2) \big], \\
    G_E^{(i)} (Q^2) &= \frac{m_0}{3(m_0 + m_i)} \big[ \big( m_i - m_0^2 -Q^2) \frac{G_1^{(i)}(Q^2)}{m_i} \\
    &+ (m_i^2 - m_{0}^2) G_2^{(i)} - 2Q^2 G_3^{(i)}(Q^2) \big], \\
    G_C^{(i)} (Q^2) &= \frac{2 m_0}{3(m_0 + m_i)} \big[ 2 m_i G_1^{(i)}(Q^2) + \frac{1}{2}(3 m_i^2 
    + m_0^2 +Q^2)G_2^{(i)}(Q^2) \\
    &+ (m_i^2 - m_{0}^2 -Q^2) G_3^{(i)}(Q^2) \big], \\
  \end{split}
\end{equation}
where $m_0$, $m_1$ and $m_2$ are the mass of the ground state $1/2$ baryons, ground and first radial excitations of decuplet baryons, respectively. Instead of studying these multipole form factors, in the present work, we analyze the magnetic dipole, $G_M^{(2)}(Q^2)$, form factor as well as the ratios of the multipole form factors $R_{SM}$ and $R_{EM}$~\cite{Braun:2005be}, which are more convenient for analyzing the experimental data
\begin{equation}
  \label{eq:14}
  \begin{split}
    R_{SM} &= - \frac{1}{4 m_2^2} \sqrt{4 m_2^2 Q^2 + (m_2^2 - Q^2 -m_N^2)^2} \frac{G_C^{(2)}(Q^2)}{G_M^{(2)}(Q^2)}, \\
     R_{EM} &= -\frac{G_E^{(2)}(Q^2)}{G_M^{(2)}{Q^2}}.
  \end{split}
\end{equation}
The main problem is the calculation of these ($G_i$ or $G_M$, $G_E$ and $G_C$) form factors. For this purpose, we implement light cone QCD sum rules. According to the LCSR methodology, we start our calculations by considering the following correlation function
\begin{equation}
  \label{eq:5}
  \Pi_{\beta \mu}(p,q) = i \int d^4x e^{i q x} \langle 0 | T \{ \eta_\beta(0)| j_\mu^{e l}(x) \} | B(p) \rangle ,
\end{equation}
where $B(p)$ is the common notation for octet $\Sigma$ and $\Xi$ baryons, $\eta_\beta$ is the interpolating current for ground or radial excitations of the  decuplet baryons, and $j_{\mu}^{\text{el}} = \sum_q e_q \bar{q} \gamma_\mu q$ is the electromagnetic current. Here $q$ denotes light quark, and $e_q$ is its electric charge. The interpolating current for decuplet baryons in a general form can be written as 
\begin{equation}
  \label{eq:6}
    \eta_\beta = N \epsilon^{abc} \big[ ({q_1^a}^T C \gamma_\beta q_2^b) q_3^c + ({q_2^a}^T C \gamma_\beta q_3^b) q_1^c  
    + ({q_3^a}^T C \gamma_\beta q_1^b) q_2^c \big],
\end{equation}
where $a,b,c$ are color indices, and $C$ is the charge conjugation operator. The quark content of the considered decuplet baryons and the value of the normalization factor is presented in Table~\ref{tab:1}.
\begin{table*}[hbt]
  \centering
  \renewcommand{\arraystretch}{1.4}
  \setlength{\tabcolsep}{7pt}
  \begin{tabular}{ccccc}
    \toprule
                       & N & $q_1$ & $q_2$ & $q_3$ \\
    \midrule
    $\Sigma^{*+}$ & $\sqrt{\frac{1}{3}}$  & $u$ & $u$ & $s$ \\ 
    $\Sigma^{*0}$ & $\sqrt{\frac{2}{3}}$  & $u$ & $d$ & $s$ \\ 
    $\Sigma^{*-}$ & $\sqrt{\frac{1}{3}}$  & $d$ & $d$ & $s$ \\ 
    $\Xi^{*0}$ & $\sqrt{\frac{1}{3}}$  & $s$ & $s$ & $u$ \\ 
    $\Xi^{*-}$ & $\sqrt{\frac{1}{3}}$  & $s$ & $s$ & $d$ \\ 
     \bottomrule
  \end{tabular}
  \caption{The quark content and the normalization factor of the decuplet baryons are presented.}
  \label{tab:1}
\end{table*}
To derive sum rules for the considered form factors, the correlation function is calculated in two different kinematical domains. In a time like domain, its expressions can be obtained by inserting the complete set of hadron states which carry the same quantum numbers as the interpolating current, and then the considered state is isolated. The phenomenological part of the correlation function can also be calculated in the deep Euclidean region $Q^2 << 0$ with the operator product expansion (OPE). In LCSR, the OPE is performed over twist. One can then obtain the relevant sum rules by matching these two representations of the correlator functions via dispersion relation. Finally, performing Borel transformation and subtracting the continuum contribution, we get the result for the considered problem.

Let us first calculate the phenomenological part of the correlation function. Inserting the hadronic states with quantum number $J^{P} = \frac{3}{2}^{+}$ and isolating the contributions of the ground and first radial excitations of  decuplet baryons we get,

\begin{equation}
  \label{eq:7}
  \Pi_{\beta \mu}(p,q) =  \sum_{i=1}^{2} \frac{\langle 0 | \eta_\beta(0)| B_i^* \rangle \langle B_i^* | j_\mu^{e l}|B(p) \rangle}{m_{i}^2 - p^{\prime^2}} + ...,  
\end{equation}
where $i=1(2)$ describe ground (first) radial excitation baryon and $...$ corresponds to the contributions of continuum and higher states. The matrix element, $\langle 0 | \eta_\beta | \beta_i^* \rangle$, is defined as;
\begin{equation}
  \label{eq:8}
  \langle 0 | \eta_\beta | B_i^* \rangle = \lambda_i u_\beta (p^\prime),
\end{equation}
where $u_\alpha(p^\prime)$ is the Rarita-Schwinger spinor, $\lambda_i$ is its residue and $p^\prime = p - q$. The second matrix element is given in~Eq.~\eqref{eq:2}. Performing summation over the spins of Rarita-Schwinger spinors using the formula,
\begin{equation}
  \label{eq:9}
  \sum_s u_\alpha^{(s)}(p^\prime) \bar{u}_\beta^{(s)} (p^\prime) =  - (\slashed{p}^\prime + m_i) \big\{ g_{\alpha \beta} - \frac{1}{3}\gamma_\alpha \gamma_\beta - \frac{2p_\alpha^{\prime} p_\beta^{\prime} }{3 m_i^2} + \frac{p_\alpha^{\prime} \gamma_\beta - p_\beta^{\prime} \gamma_\alpha}{3 m_i} \big\}
\end{equation}
for the correlation function from hadronic part, we get 
\begin{equation}
  \label{eq:8}
  \begin{split}
    \Pi_{\alpha \mu} =& -  \sum_i \frac{\lambda_i}{m_i^2 - p^{\prime 2}}(\slashed{p}^\prime + m_i) \big\{ g_{\alpha \beta} - \frac{1}{3} \gamma_\alpha \gamma_\beta - \frac{2 p_\alpha^\prime p_\beta^\prime}{3m_i^2} + \frac{p_\alpha^\prime \gamma_\beta - p_\beta^\prime \gamma_\alpha}{3 m_i} \big\} \\
    & \big\{ G_1^{i} ( - q_\beta \gamma_\mu  + g_{\beta \mu} \slashed{q}  ) +
    G_2^{i} (-q_\beta \mathcal{P}_\mu + g_{\beta \mu} q \mathcal{P}   ) +
    G_3^{i} (q_\beta q_\mu - g_{\beta \mu} q^2 ) \big\} \gamma_5 u_N(p).
\end{split}
\end{equation}

It should be noted that we face with the following drawbacks. The interpolating current $\eta_\alpha$ couples not only to spin $3/2$ state but also to spin $(1/2)^-$ one. The matrix element of the interpolating current between vacuum and spin $(1/2)^-$ state in general is determined as
\begin{equation}
  \label{eq:9}
  \langle 0 | \eta_\alpha | \frac{1}{2} (p^\prime) \rangle = (A \gamma_\alpha + B p^\prime_\alpha) u(p^\prime). 
\end{equation}
Multiplying both side to $\gamma_\alpha$ and using $\gamma^\alpha \eta_\alpha =0$, we get
\begin{equation}
  \label{eq:9}
  \langle 0 | \eta_\alpha | \frac{1}{2} (p^\prime) \rangle = \frac{B}{4}(-m \gamma_\alpha + 4 p^\prime_\alpha) u(p^\prime). 
\end{equation}
From Eq.~\eqref{eq:9}, we see that the structures proportional to $\gamma_\alpha$ or $p_\alpha^\prime$ contain the contributions of $\frac{1}{2}^{-}$ states. Hence, to take into account only the contributions of $\frac{3}{2}$ states, these structures should be removed. Considering this fact, it follows from Eq.~\eqref{eq:8}  that contribution of the spin-$3/2$ term is solely due to the term $g_{\alpha \beta}$.

There is another unpleasant situation, namely, the fact that all structures being not independent. To obtain independent structures, we need an ordering procedure. In the present work, we choose the ordering of Dirac matrices as $\gamma_\alpha \slashed{p}^\prime \slashed{q} \gamma_\mu \gamma_5$. Taking into account all the aforementioned circumstances, the hadronic part of the correlation function for describing $3/2$ decuplet baryons (radial excitations) to octet baryon transitions containing only the contributions of spin $3/2$ states can be written as follows. 
\begin{equation}
  \label{eq:10}
  \begin{split}
    \Pi_{\alpha \mu} =&  - \frac{ \lambda_1}{m_1^2 - {p^\prime}^2} (\slashed{p}^\prime + m_1) \big[ G_1^{(1)} ( - q_\alpha \gamma_\mu + g_{\alpha \mu} \slashed{q})  \\ &+
    G_2^{(1)} [ -q_\alpha (p^\prime + q/2)_\mu + q \cdot (p^\prime + q/2) g_{\alpha \mu} ]+
    G_3^{(1)} [q_\alpha q_\mu - q^2 g_{\alpha \mu}] \gamma_5 u_N(p) \big] \\& -
    \frac{\lambda_{2}}{m_2^2 - {p^\prime}^2} (\slashed{p}^\prime + m_2) \big[ G_1^{(2)} (-q_\alpha \gamma_\mu + g_{\alpha \mu} \slashed{q}) \\ &+
    G_2^{(2)} [ -q_\alpha (p^\prime + q/2)_\mu + q \cdot (p^\prime + q/2) g_{\alpha \mu} ] +
    G_3^{(2)} [q_\alpha q_\mu - q^2 g_{\alpha \mu}] \gamma_5 u_N(p) \big].
  \end{split}
\end{equation}
where $m_1(m_2)$, $G_1^{(i)}(G_2^{(i)})$ are the mass of ground (radial excitation)  $3/2$ state and form factors describing the transitions $3/2~(\text{radial excitation}) \rightarrow 1/2$, respectively.

This equation contains six form factors. Three of them correspond to the ground, and the other three correspond to radial excited state decuplet baryons to ground state octet baryon transition. To determine these six form factors, we need six independent Lorentz structures.

From Eq.~\eqref{eq:10}, we see that the correlation function can be written in terms of the six independent structures in the following way.
\begin{equation}
  \label{eq:11}
  \begin{split}
    \Pi_{\alpha \mu} &= \Pi_1 \slashed{p}^\prime \slashed{q} \gamma_5 g_{\alpha \mu} +
    \Pi_2 \slashed{q}  \gamma_5 g_{\alpha \mu} +
    \Pi_3 \slashed{p}^\prime \gamma_5 p^\prime_\mu q_\alpha +
    \Pi_4 \gamma_5 p^\prime_\mu q_\alpha +
    \Pi_5 \slashed{p}^\prime \gamma_5 q_\alpha q_\mu \\
    &+ 
    \Pi_6 \gamma_5 q_\alpha q_\mu +
    ~\text{other structures}        
  \end{split}
\end{equation}

Equating the  coefficients of the relevant Lorentz structures from Eqs.~\eqref{eq:10} and \eqref{eq:11}, we get,
\begin{equation}
  \label{eq:12}
  \begin{split}
   \Pi_1 &=  -\frac{ \lambda_1 G_1^{(1)}}{m_1^2-{p^\prime}^2} - \frac{\lambda_2 G_1^{(2)}}{m_2^2-{p^\prime}^2},  \\
   \Pi_2 &=  -\frac{\lambda_1 m_1 G_1^{(1)}}{m_1^2-{p^\prime}^2} - \frac{\lambda_2 m_2 G_1^{(2)}}{m_2^2-{p^\prime}^2},  \\
   \Pi_3 &=  \frac{\lambda_1 G_2^{(1)}}{m_1^2-{p^\prime}^2} + \frac{\lambda_2 G_2^{(2)}}{m_2^2-{p^\prime}^2},  \\
   \Pi_4 &=  \frac{\lambda_1 m_1 G_2^{(1)}}{m_1^2-{p^\prime}^2} + \frac{\lambda_2 m_2 G_2^{(2)}}{m_2^2-{p^\prime}^2},  \\
   \Pi_5 &=  \frac{\lambda_1}{m_1^2-{p^\prime}^2} [\frac{G_2^{(1)}}{2} - G_3^{(1)}] + \frac{\lambda_2}{m_2^2-{p^\prime}^2} [\frac{G_2^{(2)}}{2} - G_3^{(2)}],  \\
   \Pi_6 &=  \frac{\lambda_1 m_1}{m_1^2-{p^\prime}^2} [\frac{G_2^{(1)}}{2} - G_3^{(1)}] + \frac{\lambda_2 m_2}{m_2^2-{p^\prime}^2} [\frac{G_2^{(2)}}{2} - G_3^{(2)}].
 \end{split}
\end{equation}
Solving these equations for our goal to determine the form factors $G_1^{(2)}$, $G_2^{(2)}$, and $G_3^{(2)}$, which describe the transition of the radial excitation of decuplet baryons to the ground state octet baryons, we get
\begin{equation}
  \label{eq:13}
  \begin{split}
   - m_1 \Pi_1 + \Pi_2 &= -\frac{ \lambda_2 G_1^{(2)} }{m_2^2 - {p^\prime}^2}(m_2 - m_1), \\
   - m_1 \Pi_3 + \Pi_4 &= \frac{\lambda_2 G_2^{(2)}}{m_2^2 - {p^\prime}^2}(m_2 - m_1), \\
   - m_1 \Pi_5 + \Pi_6 &= \frac{\lambda_2}{m_2^2 - {p^\prime}^2}(m_2 - m_1) [\frac{G_2^{(2)}}{2} - G_3^{(2)}]. 
  \end{split}
\end{equation}
From this equation, it follows that to obtain the sum rules for the form factors $G_{i}^{(2)}$, the expression of the correlation function from the OPE part is needed. The OPE part of the correlation function can be obtained by inserting the expression of the interpolating current given in Eq.~\eqref{eq:6} into Eq.~\eqref{eq:5} and using the Wick theorem. As a result, one can get the expressions of invariant functions in the deep Euclidean domain ${p^\prime}^2 = (p-q)^2 << 0$. In the LCSR method, the OPE part of the correlation function for the considered problem is expressed in terms of octet baryon distribution amplitudes (DA's). The distribution amplitudes of the octet baryons appear in matrix element of three quark non-local operator between vacuum and members of the octet baryons
\begin{equation}
  \label{eq:19}
  \epsilon^{abc} \langle 0 | q_{1 \alpha}^{a} (a_1 x) q_{2 \beta}^{b}(a_2 x) q_{3 \gamma}^{c} (a_3 x) | O(p) \rangle.
\end{equation}
Using the spin and parity of the baryons as well as the Lorentz covariance, the general Lorentz composition of this matrix element can be written as
\begin{equation}
  \label{eq:20}
  \begin{split}
    4  \epsilon^{abc} \langle 0 | q_{1 \alpha}^{a} (a_1 x) q_{2 \beta}^{b}(a_2 x) q_{3 \gamma}^{c} (a_3 x) | O(p) \rangle = \sum_i \mathcal{F}_i \Gamma_{\alpha \beta}^{1 i }( \Gamma^{2 i } u(p) )_\gamma
  \end{split}
\end{equation}
where $\Gamma^{1(2)i}$ are concrete Dirac matrices, $\mathcal{F}_i = \mathcal{S}, \mathcal{P}_i, \mathcal{A}_i, \mathcal{V}_i$ and $\mathcal{T}_i$ are the DA's having no definite twists. The decomposition of Eq.~\eqref{eq:9} in terms of $\mathcal{F}_i$ is given in~\cite{Braun:2005be}. The DA's with definite twists are defined as
\begin{equation}
  \label{eq:21}
  \begin{split}
    4  \epsilon^{abc} \langle 0 | q_{1 \alpha}^{a} (a_1 x) q_{2 \beta}^{b}(a_2 x) q_{3 \gamma}^{c} (a_3 x | O(p) \rangle = \sum_i F_i \Gamma_{\alpha \beta}^{1 i }( \Gamma^{2 i } u(p) )_\gamma .
  \end{split}
\end{equation}
where $F_i = S, P_i, A_i, V_i$ and $T_i$. The relations among the two sets of DA's is given in~\cite{Braun:2005be} and for completeness presented in Appendix A. The DA's, which are the main non-perturbative ingredient, and up to twist-6 are calculated in~\cite{Braun:2005be,Bali:2015ykx,Bali:2019ecy,Wein:2015oqa,Liu:2009uc,Liu:2008yg,Braun:2006hz}.

Using the expressions of the DAs of the octet baryons, the invariant functions $\Pi_i$ is calculated straightforward. The general form of the expressions of the invariant functions can be written as
\begin{equation}
  \label{eq:22}
  \Pi_i\big( (p-q)^2,q^2) = \sum_{n=1}^{3} \int_0^1 \frac{\rho_i^{(n)} \big(x,q^2,(p-q)^2 \big)}{\big( (q-px)^2 \big)^n}
\end{equation}
where $\rho_i^n$ are spectral densities. Explicit expressions for $\rho_i^n$ for the  considered transitions are presented in Appendix B. Using the quark-hadron duality ansatz and performing Borel transformation from both side of Eq.~\eqref{eq:13} with respect to $-{p^\prime}^2 = -(p-q)^2$ to enhance the contributions of first radial excitations as well as suppressing the contributions of higher states and continuum, we arrive the desired sum rules for the transition form factors of $\gamma^* \text{octet} \rightarrow \text{first radial excitations of decuplet baryons.}$
\begin{equation}
  \label{eq:23}
  \begin{split}
    - \lambda_2 G_1^{(2)}(Q^2)(m_2 - m_1) e^{-m_2^{^2}/M^2} &= - m_1 I_1(M^2,Q^2,s_0) + I_2(Q^2,M^2,s_0), \\
      \lambda_2 G_2^{(2)}(Q^2)(m_2 - m_1) e^{-m_2^{^2}/M^2} &= - m_1 I_3(M^2,Q^2,s_0) + I_4(Q^2,M^2,s_0), \\
    \lambda_2 (\frac{G_2^{(2)}(Q^2)}{2} - G_3^{(2)}(Q^2))(m_2 - m_1) e^{-m_2^{^2}/M^2} &= -m_1 I_5(M^2,Q^2,s_0) + I_6(Q^2,M^2,s_0),
  \end{split}
\end{equation}
where the functions $I_i(,Q^2,M^2,s_0)$ are  (see \cite{Gubernari:2018wyi} and \cite{Aliev:2019ojc})
\begin{equation}
  \label{eq:20}
  \begin{split}
    I_i =& \sum_{n=1}^3  \int_{x_0}^{1} dx e^{-s/M^2} \frac{1}{(n-1)!} \frac{\rho_i^{(n)}}{x^n (M^2)^{n-1}}  \\
    &- e^{-s_0/M^2}\bigg[ \frac{(-1)^{n-1}}{(n-1)!}  \sum_{j=1}^{n-1} \frac{1}{(M^2)^{n-j-1}}\frac{1}{s^\prime}\big(\frac{d}{dx} \frac{1}{s^\prime}\big)^{j-1} \frac{\rho_i^{(n)}}{x^n}  \bigg]_{ |_{x = x_0}},  \end{split}
\end{equation}
where $s = \frac{m_{0}^2 \bar{x} x + Q^2 \bar{x}}{x}$, $\bar{x} = 1-x$, $s^\prime = \frac{ds}{dx}$, and $x_0$ is the solution of $s_0 = s$ equation in which $m_0$ is the ground state baryon mass. It follows from Eq.~\eqref{eq:23} that to determine the transition form factors, the residues of the radial excitations of decuplet baryons are needed. The mass and residues of these baryons within QCD sum rules are estimated in~\cite{Aliev:2016jnp}, and the results are: $m_{2} = 1.389~{GeV}$, $\lambda_{2} = 0.045~{GeV^3}$ (for $\Sigma^*$ baryon), $m_{2} = 1.577~{GeV}$, $\lambda_{2} = 0.045~{GeV^3}$ (for $\Xi^*$ baryon). 

Once the form factors, $G_i^{(2)}$, are calculated from the sum rule presented in Eq.~\eqref{eq:23} then using Eq.~\eqref{eq:2} we can calculate the multipole form factors $G_M^{(2)}(Q^2)$, $G_E^{(2)}(Q^2)$, and $G_C^{(2)}(Q^2)$.

\section{Numerical Analysis}
\label{sec:3}
In the present section, the numerical calculations are done for the multipole form factors as well as $R_{EM}$ and $R_{SM}$. As we already noted, the main non-perturbative inputs are DAs of the octet baryons. These DA's are presented in~\cite{Bali:2019ecy,Bali:2015ykx,Wein:2015oqa,Liu:2009uc,Liu:2008yg,Braun:2006hz}. In these works, the values of the parameters entering the expressions of DA's are also depicted. To predict the form factors reliably, the working regions of two auxiliary parameters, the Borel mass $M^2$ and the continuum threshold $s_0$, should be specified. The working regions of these parameters are determined in a standard way. The working window for $M^2$ is determined by demanding the following criteria:
\begin{enumerate}
\item The lower bound of $M^2$ is obtained by requiring that, the higher states and continuum's contributions should be by less than say $40\%$ of the total result.
\item The upper limit is determined from the condition that the higher twist conditions should be smaller than the leading twist one. Our numerical analysis leads to the conclusion that these conditions are satisfied in the regions shown below for the considered transitions.
  \begin{equation}
    \label{eq:4}
    \begin{split}
     & 1.5~GeV^2 \leq M^2 \leq~3.5~GeV^2  \hspace{1.5cm} (\text{for } \gamma^* \Sigma \rightarrow \Sigma^*) \\
     & 1.5~GeV^2 \leq M^2 \leq~3.5~GeV^2 \hspace{1.5cm} (\text{for } \gamma^* \Xi \rightarrow \Xi^*) \\
     & 2.0~GeV^2 \leq M^2 \leq~4.0~GeV^2 \hspace{1.5cm} (\text{for } \gamma^* \Xi \rightarrow \Xi^*)     
    \end{split}
  \end{equation}
\end{enumerate}
Besides, the continuum threshold $s_0$ is obtained by requiring that the mass sum rules within $10\%$ accuracy reproduce the mass of $\Sigma^*(1.727)$ and $\Xi^*(1.965)$ states. From these conditions, we get $s_0 = (2.6 \pm 0.1)^2~\rm{GeV}^2$ and $(s_0 = 2.9 \pm 0.1)^2~\rm{GeV}^2$ for $\gamma^* \Sigma \rightarrow \Sigma^*$ and $\gamma^* \Xi \rightarrow \Xi^*$ transitions, respectively.

\begin{figure}[hbt!]
  \centering
  \includegraphics[scale=0.60]{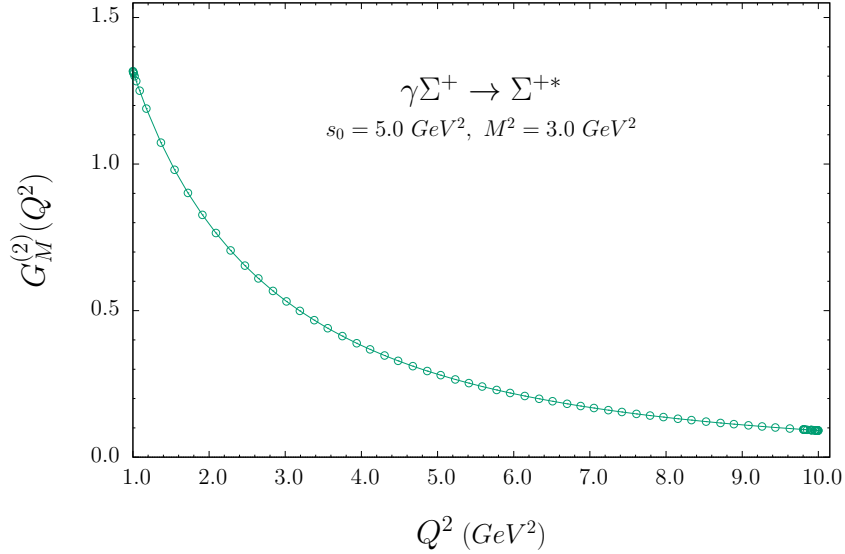}
  \caption{The dependency of the $G_M^{(2)}(Q^2)$ on $Q^2$ at a fixed values of $s_0$ and $M^2$ for $\gamma \Sigma^+ \rightarrow \Sigma^{+*}$.}
  \label{fig:1}
\end{figure}

\begin{figure}[hbt!]
  \centering
  \includegraphics[scale=0.60]{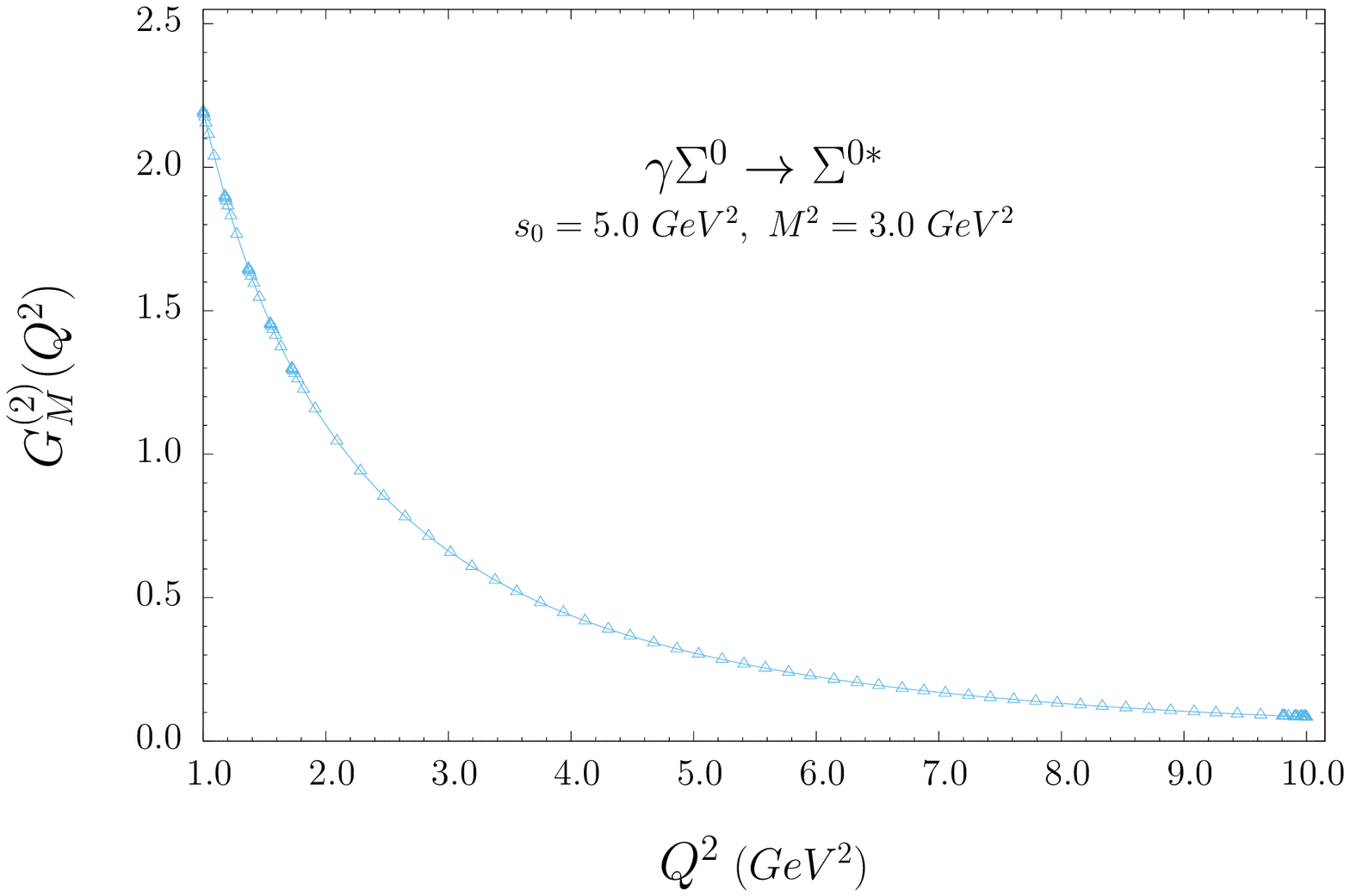}
  \caption{The same as in Fig.~\ref{fig:1}, but for $\gamma \Sigma^0 \rightarrow \Sigma^{0*}$.}
  \label{fig:2}
\end{figure}

\begin{figure}[hbt!]
  \centering
  \includegraphics[scale=0.60]{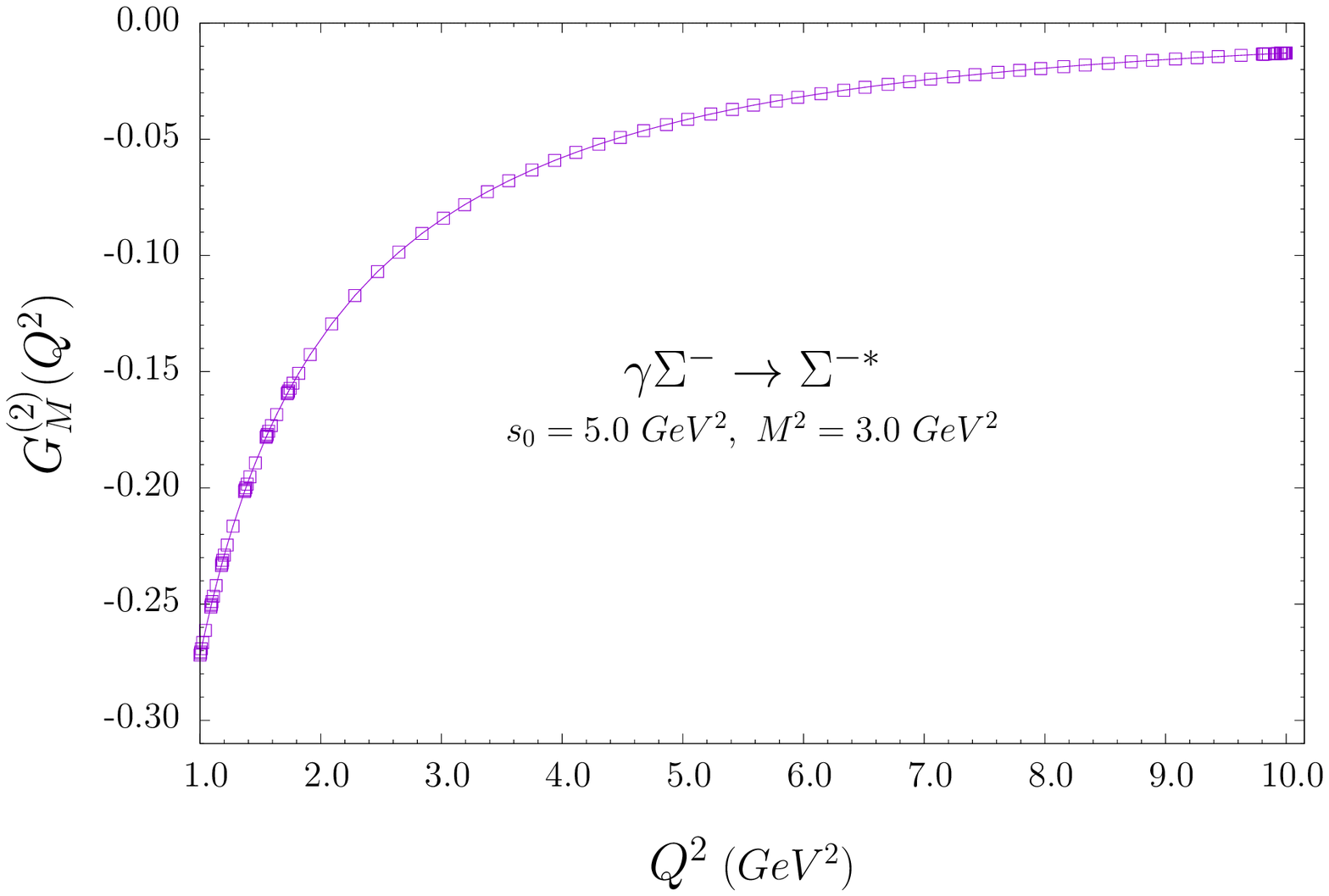}
  \caption{The same as in Fig.~\ref{fig:1}, but for $\gamma \Sigma^- \rightarrow \Sigma^{-*}$.}
  \label{fig:3}
\end{figure}

\begin{figure}[hbt!]
  \centering
  \includegraphics[scale=0.60]{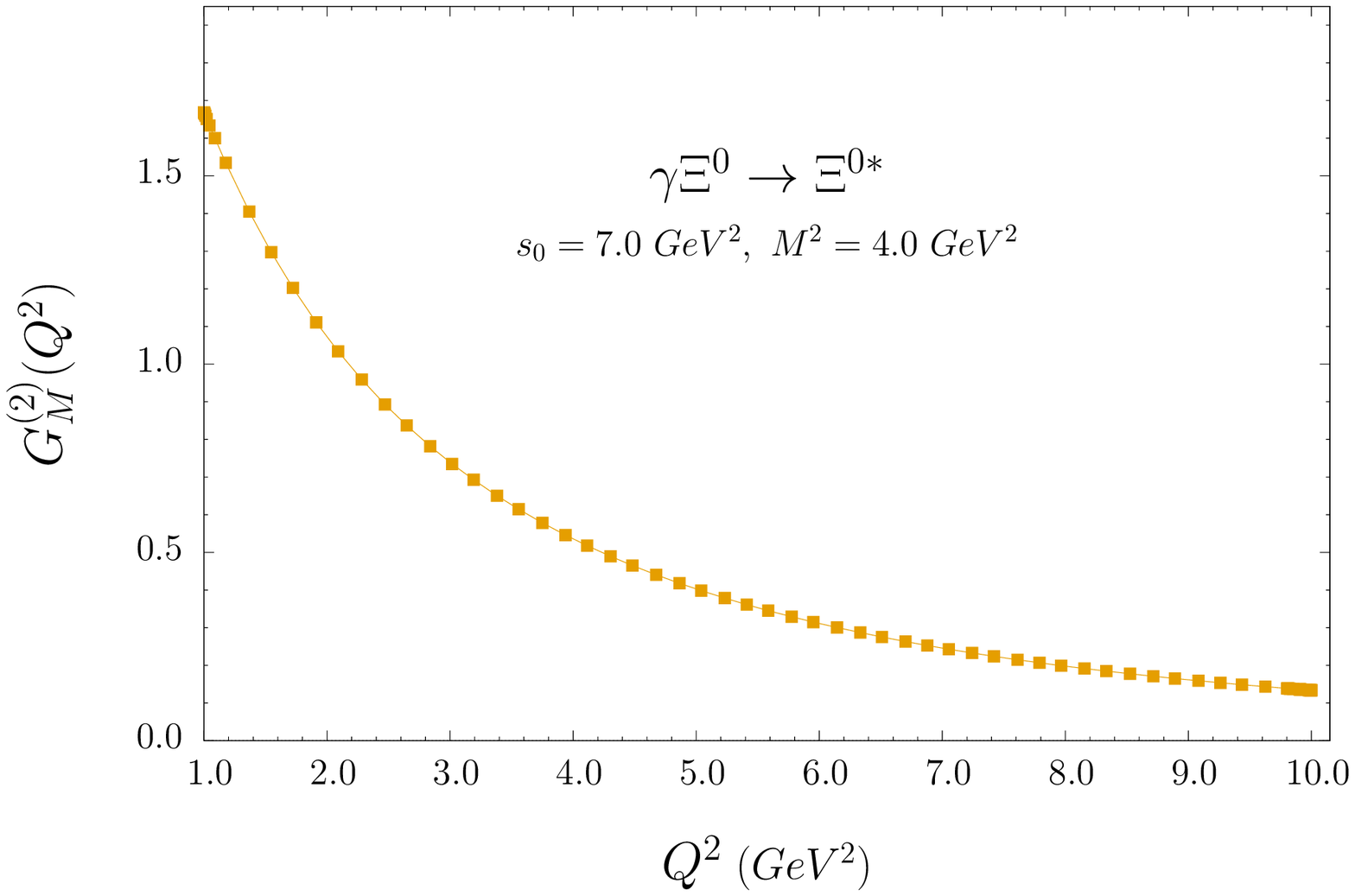}
  \caption{The same as in Fig.~\ref{fig:1}, but for  $\gamma \Xi^0 \rightarrow \Xi^{0*}$ .}
  \label{fig:4}
\end{figure}

\begin{figure}[hbt!]
  \centering
  \includegraphics[scale=0.60]{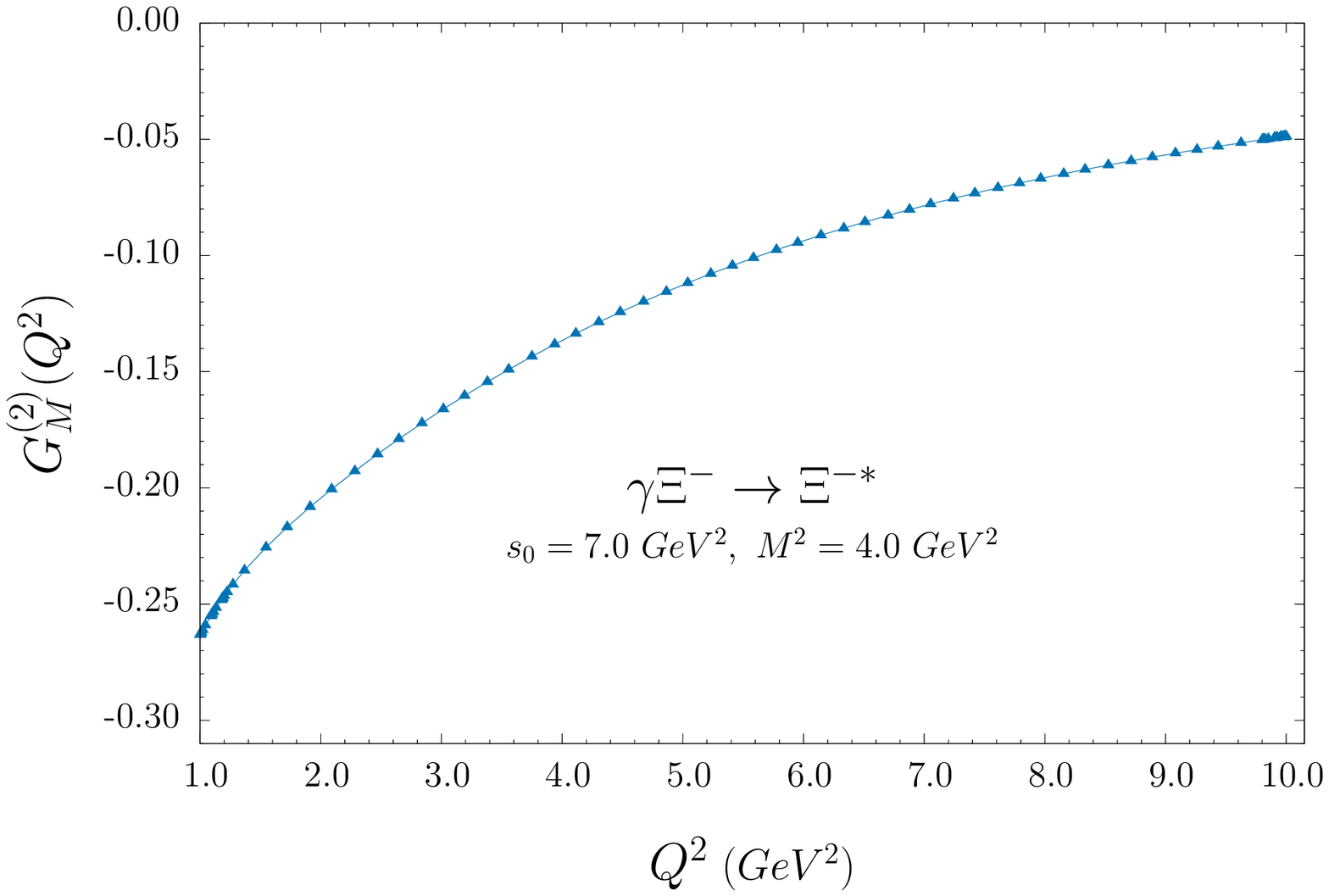}
  \caption{The same as in Fig.~\ref{fig:1}, but for  $\gamma \Xi^{-} \rightarrow \Xi^{-*}$ .}
  \label{fig:5}
\end{figure}

Having the values of all the input parameters, we can start performing numerical calculations. In~\Cref{fig:1,fig:2,fig:3,fig:4,fig:5}, we depict the dependency of the magnetic dipole form factors $G_{M}(Q^2)$ on $Q^2$ for the considered transitions at the fixed values of $s_0$ and $M^2$. From these figures, we can classify the results for the $\gamma^* \Sigma \rightarrow \Sigma^*$ and $\gamma^* \Xi \rightarrow \Xi^*$ transitions using the magnitudes of the dipole form factors $G_M$ as 
\begin{itemize}
\item \textbf{large:}  for $\gamma^* \Sigma^{+} \rightarrow \Sigma^{+ *}$, $\gamma \Sigma^0 \rightarrow \Sigma^{0*}$ and $\gamma \Xi^0 \rightarrow \Sigma^{0*}$
  \item \textbf{small:} for $\gamma^* \Sigma^-  \rightarrow \Sigma^{-*}$ and $\gamma^* \Xi^- \rightarrow \Xi^{-*}$.
\end{itemize}
This classification is similar to the ground state decuplet octet transitions, with only one exception. In the last transitions, the magnitude of $G_{M}(Q^2)$ for $\gamma^{*} \Sigma^0 \rightarrow \Sigma^{0*}$ transition is moderate, but in our case, its magnitude is comparable with $\gamma^{*} \Sigma^{+} \rightarrow \Sigma^{+*}$ and $\gamma^* \Xi^0 \rightarrow \Xi^{0*}$ transitions. This can be explained as a consequence of U-spin symmetry~\cite{Lipkin:1973rw}.

Let us consider the following ratio $T = | \frac{{G_M^{(2)}}^{\Sigma^{+}}}{{G_M^{(2)}}^{\Xi^0}} -1 |$. In the U-spin symmetry case, this factor must be equal to zero. However, from our findings, we observe that the $U-spin$ symmetry is broken about $15\%$. Similar results can be inferred for the $T$-factor  of the $\gamma^* \Sigma^- \rightarrow \Sigma^{-*}$ and $\gamma^* \Xi^- \rightarrow \Xi^{-*}$ transitions.

\begin{figure}[hbt!]
  \centering
  \includegraphics[scale=0.60]{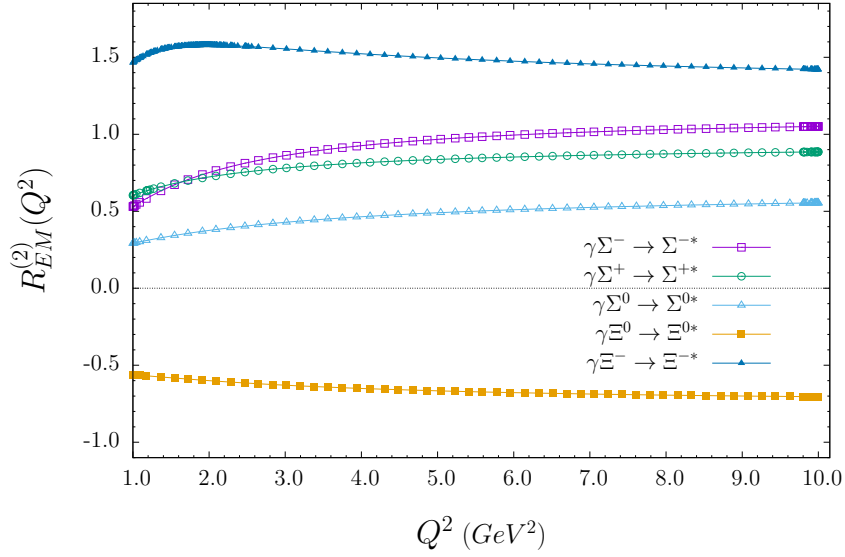}
  \caption{The dependency of the $R_{EM}(Q^2)$ on $Q^2$ at $s_0 = 5~\rm{GeV^2}$ and $M^2 = 3~\rm{GeV^2}$ for $\gamma \Sigma^+ \rightarrow \Sigma^{+*}$, $\gamma \Sigma^- \rightarrow \Sigma^{-*}$, and $\gamma \Sigma^0 \rightarrow \Sigma^{0*}$ and $s_0 = 7~\rm{GeV^2}$ and $M^2 = 4~\rm{GeV^2}$ for  $\gamma \Xi^0 \rightarrow \Xi^{0*}$ and  $\gamma \Xi^- \rightarrow \Xi^{-*}$}
  \label{fig:6}
\end{figure}

\begin{figure}[hbt!]
  \centering
  \includegraphics[scale=0.60]{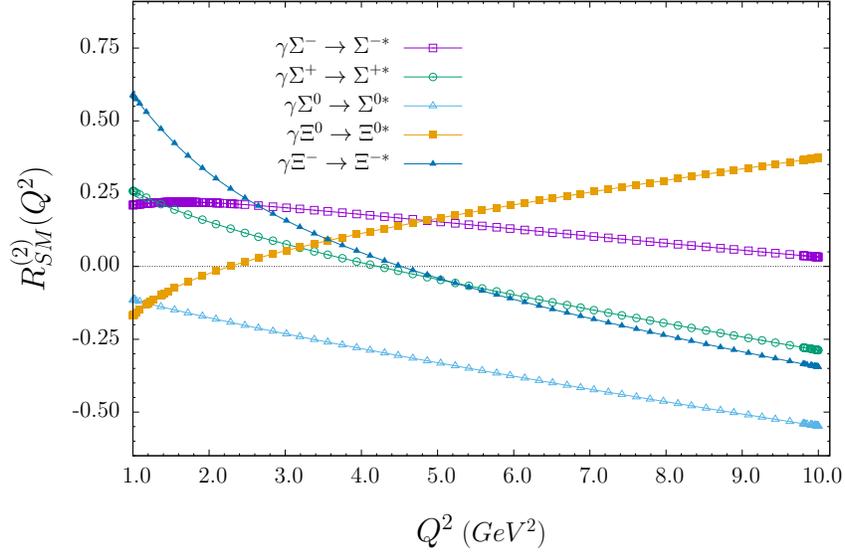}
  \caption{Same as in Fig.~\ref{fig:6} but for $R_{SM}$.} 
  \label{fig:7}
\end{figure}

Instead of studying the behavior of $G_E^{(2)}(Q^2)$ and $G_C^{(2)}(Q^2)$ on $Q^2$, we study the dependencies of $R_{EM}$ and $R_{SM}$ on $Q^2$ since these quantities are related with $G_E^{(2)}(Q^2)$ and $G_C^{(2)}(Q^2)$ (see Eq.~\eqref{eq:14}). The dependencies of these quantities on $Q^2$ are presented in Figs.~\ref{fig:6} and \ref{fig:7}, respectively. From these figures, we get the following domains of variation of $R_{EM}$ and $R_{SM}$:
\begin{table*}[hbt]
  \centering
  \renewcommand{\arraystretch}{1.4}
  \setlength{\tabcolsep}{7pt}
  \begin{tabular}{ccc}
    \toprule
                       & $R_{EM}(Q^2)$ & $R_{SM}(Q^2)$\\
    \midrule
    $\gamma^* \Sigma^0 \rightarrow \Sigma^{0*}$ & $ 0.28 \div 0.55$  & $ -0.1 \div -0.5$  \\ 
    $\gamma^* \Sigma^- \rightarrow \Sigma^{-*}$ & $ 0.4 \div 0.8$  & $ 0.15 \div 0.30$  \\ 
    $\gamma^* \Sigma^+ \rightarrow \Sigma^{+*}$ & $ 0.50 \div 0.85$  & $ -0.25 \div 0.25$  \\ 
    $\gamma^* \Xi^0 \rightarrow \Xi^{0*}$ & $ -0.50 \div -0.75$  & $ -0.2\div 0.4$  \\ 
    $\gamma^* \Xi^- \rightarrow \Xi^{-*}$ & $ 1.3 \div 1.4$  & $ -0.2 \div 0.5$  \\ 

    \bottomrule
  \end{tabular}
  \caption{The obtained results of $R_{EM}$ and $R_{SM}$ for the considered transitions are shown with respect to the variation of $Q^2$.}
  \label{tab:2}
\end{table*}
For comparison, we also present the results of these quantities for the ground state decuplet to ground state octet baryons in Table~\ref{tab:3}.
\begin{table*}[hbt]
  \centering
  \renewcommand{\arraystretch}{1.4}
  \setlength{\tabcolsep}{7pt}
  \begin{tabular}{ccc}
    \toprule
                       & $R_{EM}(Q^2)$ & $R_{SM}(Q^2)$\\
    \midrule
    $\gamma^* \Sigma^0 \rightarrow \Sigma^{0*}$ & $ 0.21 \div 0.2$  & $ -0.3 \div -0.5$  \\ 
    $\gamma^* \Sigma^- \rightarrow \Sigma^{-*}$ & $ -0.35 \div 0.6$  & practically zero  \\ 
    $\gamma^* \Sigma^+ \rightarrow \Sigma^{+*}$ & $ 0.1 \div 0.2$  & $ -0.25 \div -0.3$  \\ 
    $\gamma^* \Xi^0 \rightarrow \Xi^{0*}$ & $ -0.1 \div 0$  & $ -0.35\div -0.4$  \\ 
    $\gamma^* \Xi^- \rightarrow \Xi^{-*}$ & $ -0.5 \div 0.8$  & $ -0.1 \div 0$  \\ 
    \bottomrule
  \end{tabular}
  \caption{The results for the ground state decuplet to ground state octet baryons are shown~~\cite{Aliev:2013jta}.}
  \label{tab:3}
\end{table*}
From Table~\ref{tab:2}, it follows that the values of $R_{EM}(Q^2)$ for $\gamma^* \Sigma^{\pm} \rightarrow \Sigma^{\pm *}$ and $\gamma^* \Xi^{0} \rightarrow \Xi^{0*}$ in magnitude very close to each other as a consequence of $U-spin$ symmetry. The large value of $R_{EM}$ for $\gamma^* \Xi^- \rightarrow \Xi^{-*}$ channel is due to the large value of $G_E^{(2)}(Q^2)$. The magnitude of $R_{SM}(Q^2)$ for the channels $\gamma^* \Sigma^- \rightarrow \Sigma^{-*}$, $\gamma^* \Sigma^+ \rightarrow \Sigma^{+*}$ and $\gamma^* \Xi^0 \rightarrow \Xi^{0*}$ are very close to each other likewise for the magnitudes.

Similar situation takes place for $\gamma^* \Sigma^0 \rightarrow \Sigma^{0*}$ and $\gamma^{*} \Xi^- \rightarrow\Xi^{-*}$ channels too. Comparing the results in Table~\ref{tab:2} and \ref{tab:3}, we observe that the values of $R_{SM}$ for the first radial decuplet to octet baryons are close to the corresponding results for the ground state decuplet to octet baryon transitions except the results for $\gamma^* \Sigma^- \rightarrow \Sigma^{-*}$ and $\gamma^* \Xi^- \rightarrow \Xi^{-*}$. On the other hand, for $R_{EM}(Q^2)$, we find that the results are close only for $\gamma^* \Sigma^- \rightarrow \Sigma^{-*}$ channels. 

\section{Conclusion}
In this work, we studied the electromagnetic transitions among the first radial excitations of decuplet baryons to ground state baryons using the light-cone QCD sum rules. 
The magnetic dipole form factors $G_M^{(2)}(Q^2)$ as well as the ratios $R_{EM} = - \frac{G_E^{(2)}(Q^2)}{G_M^{(2)}(Q^2)}$ and $R_{SM} = - \frac{1}{4 m_1^2} \sqrt{4 m_1^2 Q^2 + (m_1^2 - Q^2 -m_N^2)^2} \frac{G_C^{(2)}(Q^2)}{G_M^{(2)}(Q^2)}$ are calculated when $Q^2$ varies in the domain $1~\rm{GeV^2} \leq Q^2 \leq 10~\rm{GeV^2}$. We observe that the results of the ${G_{M}^{(2)}}(Q^2)$ for the transitions $\gamma^* \Sigma^+ \rightarrow \Sigma^{*+}$, $\gamma^* \Xi^- \rightarrow \Xi^{-*}$ are very close to each other as a result of the SU(3) symmetry. We also find that the results for $R_{SM}(Q^2)$ for the first radial decuplet excitation state to octet baryons are close to each other  except the results for the $\gamma \Sigma^{-} \rightarrow \Sigma^{-*}$ and $\gamma^{*} \Xi^{-} \rightarrow \Xi^{-*}$ transitions. For the quantity $R_{EM}$, we see that the result only for $\gamma^{*} \Sigma^{-} \rightarrow \Sigma^{-*}$ and more or less for $\gamma^* \Xi^{-} \rightarrow \Xi^{-*}$ channels to considered transitions are close to each other.

The results of the multipole form factors for the $\gamma^* B \rightarrow B^{*}$ octet to the first radial excitation of decuplet baryons transition can be very useful and play a critical role for studies of the properties of the excited baryon spectrum and their properties in future experiments.

\clearpage


\bibliographystyle{apsrev4-1}
\bibliography{multipole.bib}


\section*{Appendix}
\label{sec:appendix}
\setcounter{equation}{0}

Relations between two sets of octet baryon distribution amplitudes are given below (see~\cite{Braun:2006hz}).

\bea
\label{nolabel}
\begin{array}{ll}
{\cal S}_1 = S_1~,& (2 P \mcdot x) \, {\cal S}_2 = S_1 - S_2~, \nnb \\
{\cal P}_1 = P_1~,& (2 P \mcdot x) \, {\cal P}_2 = P_2 - P_1~, \nnb \\
{\cal V}_1 = V_1~,& (2 P \mcdot x) \, {\cal V}_2 = V_1 - V_2 - V_3~, \nnb \\
2{\cal V}_3 = V_3~,& (4 P \mcdot x) \, {\cal V}_4 =
- 2 V_1 + V_3 + V_4 + 2 V_5~,\nnb \\
(4 P\mcdot x) \, {\cal V}_5 = V_4 - V_3~,&
(2 P \mcdot x)^2 \, {\cal V}_6 = - V_1 + V_2 + V_3 + V_4 + V_5 - V_6~, \nnb \\
{\cal A}_1 = A_1~,& (2 P \mcdot x) \, {\cal A}_2 = - A_1 + A_2 - A_3~, \nnb \\
2 {\cal A}_3 = A_3~,&
(4 P \mcdot x) \, {\cal A}_4 = - 2 A_1 - A_3 - A_4 + 2 A_5~, \nnb \\
(4 P \mcdot x) \, {\cal A}_5 = A_3 - A_4~,&
(2 P \mcdot x)^2 \, {\cal A}_6 = A_1 - A_2 + A_3 + A_4 - A_5 + A_6~, \nnb \\
{\cal T}_1 = T_1~, & (2 P \mcdot x) \, {\cal T}_2 = T_1 + T_2 - 2T_3~, \nnb \\
2 {\cal T}_3 = T_7~,& (2 P \mcdot x) \, {\cal T}_4 = T_1 - T_2 - 2 T_7~, \nnb \\
(2 P\mcdot x) \, {\cal T}_5 = - T_1 + T_5 + 2 T_8~,&
(2 P \mcdot x)^2 \, {\cal T}_6 = 2 T_2 - 2 T_3 - 2 T_4 + 2 T_5 + 2 T_7 + 2 T_8~, \nnb \\
(4P \mcdot x) \, {\cal T}_7 = T_7 - T_8~, &
(2 P\mcdot x)^2 \, {\cal T}_8 = - T_1 + T_2 + T_5 - T_6 + 2 T_7 + 2 T_8~. \nnb
\end{array}
\eea
Explicit expressions of DAs ${\cal S}_i$, ${\cal P}_i$, ${\cal A}_i$, ${\cal V}_i$
and ${\cal T}_i$ at the leading order of conformal spin expansion
can be found in~\cite{Braun:2005be,Bali:2015ykx,Bali:2019ecy,Wein:2015oqa,Liu:2009uc,Liu:2008yg,Braun:2006hz}.

\section*{Appendix B}  
\setcounter{equation}{0}

In this appendix, we present the expressions for the functions
$\rho_1^{(i)}$, $\rho_2^{(i)}$ and $\rho_3^{(i)}$
which appear in the sum rules for $G_1^{(2)}(Q^2)$,
$G_2^{(2)}(Q^2)$, and $\ds{G_2^{(2)}(Q^2)\over 2} - G_3^{(2)}(Q^2)$, for the $\gamma^\ast
\Sigma^+ \to \Sigma^{\ast +}$ transition.

\section*{Functions $\rho_i^{(n)}$ for the form factor $G_1^{(2)}$}

\begin{equation}
  \label{eq:100}
  \begin{split}
\rho_{1}^{(3)} (x)  &= 0 \\
\rho_{2}^{(3)} (x) &=  \frac{8 (1-x)}{x} e_{q_2}   m_{0}^2 m_{q_2} (x^2  m_{0}^2+Q^{2}) \, \widetilde{\!\widetilde{B}}_6     \\
\rho_{1}^{(2)} (x) &=  -4 e_{q_3}   m_{0} ( m_{0} \widehat{\!\widehat{B}}_6-2 m_{q_3} \widehat{B}_4) + 8 e_{q_2}   m_{0} m_{q_2} \widetilde{B}_2 \\
&-8 e_{q_2}   m_{0}^2 \int_0^{\bar{x}} dx_1 
 ( {T_1}^{M}-{A_1}^{M}) ( x_1, x, 1-x_1 - x) \\
 &+8 e_{q_3}   m_{0}^2 \int_0^{\bar{x}} dx_1 
{T_1}^{M} ( x_1, 1-x_1 - x, x) \\
\rho_{2}^{(2)} (x)  &=  - \frac{4  m_{0}}{x} \Big\{
 -e_{q_1}  \Big[(x-1) (x^2  m_{0}^2+Q^{2}) \check{C}_2 + 2 x (x+1) \check{D}_2 \Big] \\
 &+e_{q_2}  \Big[ x^3  m_{0}^2 \widetilde{B}_2+x^3  m_{0}^2 \widetilde{B}_4+(x-1) (x^2  m_{0}^2+Q^{2}) \widetilde{D}_2-(x-1) (x^2  m_{0}^2+Q^{2}) \widetilde{C}_2 \\
 &-x^2  m_{0}^2 \widetilde{B}_2-x^2  m_{0}^2 \widetilde{B}_4-2 x^2  m_{0} m_{q_2} \widetilde{H}_1+2 x^2  m_{0} m_{q_2} \widetilde{B}_5+2 x^2  m_{0} m_{q_2} \widetilde{B}_7 \\
 &+2 (x-1) x  m_{0}^2 \, \widetilde{\!\widetilde{B}}_8+2 x  m_{0} m_{q_2} \widetilde{H}_1-2 x  m_{0} m_{q_2} \widetilde{B}_5-2 x  m_{0} m_{q_2} \widetilde{B}_7-x  m_{0} m_{q_2} \, \widetilde{\!\widetilde{B}}_6 \\
 &+x Q^{2} \widetilde{B}_2+x Q^{2} \widetilde{B}_4-2  m_{0} m_{q_2} \, \widetilde{\!\widetilde{B}}_6-Q^{2} \widetilde{B}_2-Q^{2} \widetilde{B}_4 \Big]  \\
 &+x e_{q_3}   m_{0} \Big[ (x-1) \Big( m_{0} ( \, \widehat{\!\widehat{D}}_6-2 \; \widehat{\!\widehat{C}}_6) 
 +m_{q_3} (\widehat{D}_5-2 \widehat{C}_5+2 \widehat{B}_5+4 \widehat{B}_7) \Big)  + m_{q_3} \, \widehat{\!\widehat{B}}_6 \Big]
 \Big\} \\
\rho_{1}^{(1)} (x) &= -e_{q_2} \int_0^{\bar{x}} dx_1 
  (8 {B_1}-8 {D_1}) ( x_1, x, 1-x_1 - x)
+8 e_{q_3} \int_0^{\bar{x}} dx_1 
{B_1} ( x_1, 1-x_1 - x, x)  \\
\rho_{2}^{(1)} (x) &= \frac{4 m_{0}}{x} \Big[ e_{q_2}    (\widetilde{D}_2-\widetilde{C}_2+\widetilde{B}_2+\widetilde{B}_4)- e_{q_1}    (x \check{D}_2+\check{C}_2) \Big] \\
&+ 4 (x-1) e_{q_1}   m_{0} \int_0^{\bar{x}} dx_3 ({C_3}-{D_3}) ( x , 1-x - x_3,x_3) \\
    &-4e_{q_2}   (x-1)  m_{0} \int_0^{\bar{x}} dx_1  \Big[ ({D_3}-{C_3}  +2 {P_1}-2 {S_1})-8 m_{q_2} {B_1} \Big] ( x_1, x, 1-x_1 - x) \\
    &+4 e_{q_3}   (x-1)  m_{0} \int_0^{\bar{x}} dx_1 \Big[ ({D_3}-2 {C_3})-8 m_{q_3} {B_1} \Big] ( x_1, 1-x_1 - x, x)
  \end{split}
\end{equation}

\section*{Functions $\rho_i$ for the form factor $G_2^{(2)}$}

\begin{equation}
  \label{eq:101}
  \begin{split}
    \rho_{3}^{(3)} (x)  &= 64 (x-1) x^2 e_{q_1}   m_{0}^3 \, \check{\!\check{C}}_6 +  16 x e_{q_2}   m_{0}^2 \Big[ 4 (x-1) x  m_{0}  (\, \widetilde{\!\widetilde{C}}_6-2 \, \widetilde{\!\widetilde{B}}_8)-m_{q_2} \, \widetilde{\!\widetilde{B}}_6 \Big] \\
    &+16 x e_{q_3}   m_{0}^2 \Big[ m_{q_3} \, \widehat{\!\widehat{B}}_6-2 (x-1) x  m_{0} ( \, \widehat{\!\widehat{D}}_6-2 \; \widehat{\!\widehat{C}}_6+2 \, \widehat{\!\widehat{B}}_8) \Big]    \\
    \rho_{4}^{(3)} (x) &= 32 (x-1) e_{q_2}   m_{0}^2 \Big[ (x^2  m_{0}^2+2 x Q^{2}-Q^{2}) \, \widetilde{\!\widetilde{B}}_6+ x  m_{0} m_{q_2} \, \widetilde{\!\widetilde{B}}_8 \Big] \\
    &+16 (x-1) e_{q_3}   m_{0}^2 \Big[ 2 x (x  m_{0}^2+Q^{2} ) \, \widehat{\!\widehat{B}}_6-x  m_{0} m_{q_3} (\, \widehat{\!\widehat{D}}_6+2 \; \widehat{\!\widehat{C}}_6)\Big] \\
    \rho_{3}^{(2)} (x) &= -8 (1-2 x) x e_{q_1}   m_{0} \check{C}_2 + 8 x e_{q_2}   m_{0} \Big[ (2 x-1) \widetilde{C}_2+2 (1-2 x) \widetilde{B}_4-\widetilde{D}_2-2 \widetilde{B}_2 \Big] \\
    & -8 x e_{q_3}   m_{0} \Big[ x \widehat{D}_2-2 x \widehat{C}_2+2 (x-1) \widehat{B}_4 \Big] \\
    \rho_{4}^{(2)}(x)  &= -8 (x-1) x e_{q_1}   m_{0}^2 (\check{D}_5-\check{C}_4)-8 e_{q_2}   m_{0} \Big\{(x-1) x  m_{0} \Big[ \widetilde{D}_5-\widetilde{C}_4-2 (\widetilde{H}_1+\widetilde{E}_1-\widetilde{B}_5)\Big]\\
    &+(4 x-3)  m_{0} \, \widetilde{\!\widetilde{B}}_6+2 x m_{q_2} \widetilde{B}_4+m_{q_2} (\widetilde{B}_2-\widetilde{B}_4) \Big\}+8 x e_{q_3}   m_{0} \Big[(x-1)  m_{0} (\widehat{D}_5+2 \widehat{C}_5-2 \widehat{B}_5) \\
    &+m_{q_3} (\widehat{D}_2+2 \widehat{C}_2)\Big]
    + 8 (2 x-1) e_{q_1}   m_{0}^2 \int_0^{\bar{x}} dx_3 {V_1}^{M} ( x, 1-x - x_3,x_3) \\
    & -8 e_{q_2}   m_{0}^2 \int_0^{\bar{x}} dx_1  \Big[{A_1}^{M}+(1-2 x) {V_1}^{M}+2 (x-1) {T_1}^{M}  \Big] ( x_1, x, 1-x_1 - x) \\
   & -16 x e_{q_3}   m_{0}^2 \int_0^{\bar{x}} dx_1 
   {T_1}^{M} ( x_1, 1-x_1 - x, x) \\
   \rho_{3}^{(1)}(x)  &= 0 \\
    \rho_{4}^{(1)}(x)  &= 8 (2 x-1) e_{q_1}  \int_0^{\bar{x}} dx_3
{C_1}( x, 1-x - x_3,x_3) \\
 & -8 e_{q_2} \int_0^{\bar{x}} dx_1  \Big[{D_1}-(2 x-1) {C_1} + 2 (x-1) {B_1} \Big] ( x_1, x, 1-x_1 - x) \\
 &-
    16 x e_{q_3} \int_0^{\bar{x}} dx_1  {B_1} ( x_1, 1-x_1 - x, x)      
  \end{split}
\end{equation}

\section*{Functions $\rho_i$ for the form factor $\frac{G_2^{(2)}}{2} - G_3^{(2)}$}

\begin{equation}
  \label{eq:102}
  \begin{split}
    \rho_{5}^{(3)} (x)  &= -64 (x-1)^2 x e_{q_1}   m_{0}^3 \, \check{\!\check{C}}_6 - 16 (x-1) e_{q_2}   m_{0}^2 \Big[ 4 (x-1) x  m_{0} ( \, \widetilde{\!\widetilde{C}}_6-2 \, \widetilde{\!\widetilde{B}}_8)-2 m_{q_2} \, \widetilde{\!\widetilde{B}}_6 \Big] \\
    &- 16 (x-1) e_{q_3}   m_{0}^2 \Big[ 2 (x-1) x  m_{0} (2 {C_6}-\widehat{\!\widehat{D}}_6- 2 \, \widehat{\!\widehat{B}}_8)+m_{q_3} \, \widehat{\!\widehat{B}}_6\Big] \\
    \rho_{6}^{(3)} (x)  &=  \frac{- 32 (x-1)^2}{x} e_{q_2}   m_{0}^2 \Big[ (x^2  m_{0}^2+2 x Q^{2}-Q^{2}) \, \widetilde{\!\widetilde{B}}_6+x  m_{0} m_{q_2} \, \widetilde{\!\widetilde{B}}_8 \Big]\\
    & -\frac{16 (x-1)^2}{x} e_{q_3}   m_{0}^2 \Big[2 x (x  m_{0}^2+Q^{2}) \, \widehat{\!\widehat{B}}_6-x  m_{0} m_{q_3} (2 {C_6}+\widehat{\!\widehat{D}}_6)\Big] \\
    \rho_{5}^{(2)} (x)  &= - 16 (x-1) x e_{q_1}   m_{0} \check{C}_2 + 16 (x-1) e_{q_2}   m_{0} (-x \widetilde{C}_2+2 x \widetilde{B}_4+\widetilde{D}_2+\widetilde{B}_2)\\
    &+ 8 (x-1) e_{q_3}   m_{0} \Big[x \widehat{D}_2-2 x \widehat{C}_2+2 (x-1) \widehat{B}_4\Big]  \\
    \rho_{6}^{(2)} (x)  &= 8 (x-1)^2 e_{q_1}   m_{0}^2 (\check{D}_5-\check{C}_4) 
    + \frac{8 (x-1)}{x} e_{q_2}   m_{0} \Big\{ (x-1) x  m_{0} \Big[ \widetilde{D}_5-\widetilde{C}_4-2 (\widetilde{H}_1+\widetilde{E}_1-\widetilde{B}_5)\Big] \\
    &+4 (x-1)  m_{0}\, \widetilde{\!\widetilde{B}}_6+2 x m_{q_2} \widetilde{B}_4 \Big\} \\
    &-8 (x-1) e_{q_3}   m_{0} \Big[(x-1)  m_{0} (\widehat{D}_5+2 \widehat{C}_5-2 \widehat{B}_5)+m_{q_3} (\widehat{D}_2+2  \widehat{C}_2)\Big] \\
    &-16 (x-1) e_{q_1}   m_{0}^2 \int_0^{\bar{x}} dx_3 {V_1}^{M} ( x, 1-x - x_3,x_3) \\
    &-16 (x-1) e_{q_2}   m_{0}^2 \int_0^{\bar{x}} dx_1  ({V_1}^{M}-{T_1}^{M})( x_1, x, 1-x_1 - x) \\
    &+16 (x-1) e_{q_3}   m_{0}^2 \int_0^{\bar{x}} dx_ {T_1}^{M}  ( x_1, 1-x_1 - x, x) \\
    \rho_{5}^{(1)} (x)  &= 0 \\
    \rho_{6}^{(1)}(x)  &= -16 (x-1) e_{q_1} \int_0^{\bar{x}} dx_3  {C_1} ( x, 1-x - x_3,x_3) \\
    &-16 (x-1) e_{q_2} \int_0^{\bar{x}} dx_1   ({C_1}-{B_1}) ( x_1, x, 1-x_1 - x) \\
    &+16 (x-1) e_{q_3} \int_0^{\bar{x}} dx_1  {B_1} ( x_1, 1-x_1 - x, x)
  \end{split}
\end{equation}

where $q_1=u$, $q_2=u$, and $q_3=d$, respectively.

In the above expressions for $\rho_2$, $\rho_4$, and $\rho_6$
the functions ${\cal F}(x_i)$ are defined in the following way:

\bea
\label{nolabel}
\check{\cal F}(x_1) \es \int_1^{x_1}\!\!dx_1^{'}\int_0^{1-
x^{'}_{1}}\!\!dx_3\,
{\cal F}(x_1^{'},1-x_1^{'}-x_3,x_3)~, \nnb \\
\check{\!\!\!\;\check{{\cal F}}}(x_1) \es
\int_1^{x_1}\!\!dx_1^{'}\int_1^{x^{'}_{1}}\!\!dx_1^{''}
\int_0^{1- x^{''}_{1}}\!\!dx_3\,
{\cal F}(x_1^{''},1-x_1^{''}-x_3,x_3)~, \nnb \\
\widetilde{\cal F}(x_2) \es \int_1^{x_2}\!\!dx_2^{'}\int_0^{1-
x^{'}_{2}}\!\!dx_1\,
{\cal F}(x_1,x_2^{'},1-x_1-x_2^{'})~, \nnb \\
\widetilde{\!\widetilde{\cal F}}(x_2) \es
\int_1^{x_2}\!\!dx_2^{'}\int_1^{x^{'}_{2}}\!\!dx_2^{''}
\int_0^{1- x^{''}_{2}}\!\!dx_1\,
{\cal F}(x_1,x_2^{''},1-x_1-x_2^{''})~, \nnb \\
\widehat{\cal F}(x_3) \es \int_1^{x_3}\!\!dx_3^{'}\int_0^{1-
x^{'}_{3}}\!\!dx_1\,
{\cal F}(x_1,1-x_1-x_3^{'},x_3^{'})~, \nnb \\
\widehat{\!\widehat{\cal F}}(x_3) \es
\int_1^{x_3}\!\!dx_3^{'}\int_1^{x^{'}_{3}}\!\!dx_3^{''}
\int_0^{1- x^{''}_{3}}\!\!dx_1\,
{\cal F}(x_1,1-x_1-x_3^{''},x_3^{''})~.\nnb
\eea
Definitions of the functions $B_i$, $C_i$, $D_i$, $E_1$ and $H_1$
that appear in the expressions for $\rho_i^{(n)}(x)$ are given as follows:

\bea
\label{nolabel}
B_2 \es T_1+T_2-2 T_3~, \nnb \\
B_4 \es T_1-T_2-2 T_7~, \nnb \\
B_5 \es - T_1+T_5+2 T_8~, \nnb \\
B_6 \es 2 T_1-2 T_3-2 T_4+2 T_5+2 T_7+2 T_8~, \nnb \\
B_7 \es T_7-T_8~, \nnb \\
B_8 \es  -T_1+T_2+T_5-T_6+2 T_7+2T_8~, \nnb \\
C_2 \es V_1-V_2-V_3~, \nnb \\
C_4 \es -2V_1+V_3+V_4+2V_5~, \nnb \\
C_5 \es V_4-V_3~, \nnb \\
C_6 \es -V_1+V_2+V_3+V_4+V_5-V_6~, \nnb \\
D_2 \es -A_1+A_2-A_3~, \nnb \\
D_4 \es -2A_1-A_3-A_4+2A_5~, \nnb \\
D_5 \es A_3-A_4~, \nnb \\
D_6 \es A_1-A_2+A_3+A_4-A_5+A_6~, \nnb \\
E_1 \es S_1-S_2~, \nnb \\
H_1 \es P_2-P_1~. \nnb
\eea

The expressions of the functions $V_i$, $A_i$, $T_i$, $S_i$ and $P_i$ are presented in Appendix A.


\end{document}